\documentclass[iop]{emulateapj}

\usepackage{times,natbib,graphicx,amsmath}

\shorttitle{A Hard Look at 4U 1636-53, GX 17+2, and 4U 1705-44}
\shortauthors{Ludlam et al.}

\begin{document}

\title{A Hard Look at the Neutron Stars and Accretion Disks in 4U 1636-53, GX 17+2, and 4U 1705-44 with $\emph{NuSTAR}$}
\author{R. M. Ludlam\altaffilmark{1},
J. M. Miller\altaffilmark{1}, 
M. Bachetti\altaffilmark{2}, 
D. Barret\altaffilmark{3,4},
A. C. Bostrom\altaffilmark{1}, 
E. M. Cackett\altaffilmark{5},  
N. Degenaar\altaffilmark{6,7},  
T. Di Salvo\altaffilmark{8}, 
L. Natalucci\altaffilmark{9}, 
J. A. Tomsick\altaffilmark{10}, 
F. Paerels\altaffilmark{11},
M. L. Parker\altaffilmark{6}}
\altaffiltext{1}{Department of Astronomy, University of Michigan, 1085 South University Ave, Ann Arbor, MI 48109-1107, USA}
\altaffiltext{2}{INAF/Osservatorio Astronomico di Cagliari, via della Scienza 5, I-09047 Selargius (CA), Italy}
\altaffiltext{3}{Universit de Toulouse; UPS-OMP; IRAP; Toulouse, France}
\altaffiltext{4}{CNRS; Institut de Recherche en Astrophysique et Plantologie; 9 Av. colonel Roche, BP 44346, F-31028 Toulouse cedex 4, France}
\altaffiltext{5}{Department of Physics \& Astronomy, Wayne State University, 666 W. Hancock St., Detroit, MI 48201, USA}
\altaffiltext{6}{Institute of Astronomy, Madingley Road, Cambridge CB3 0HA, UK}
\altaffiltext{7}{Anton Pannekoek Institute for Astronomy, University of Amsterdam, Pastbus 94249, 1090 GE Amsterdam, The Netherlands}
\altaffiltext{8}{Dipartimento di Fisica e chimica, Universit\'{a} degli Studi di Palermo, via Archirafi 36, 90123, Palermo, Italy}
\altaffiltext{9}{Istituto Nazionale di Astrofisica, INAF-IAPS, via del Fosso del Cavaliere, 00133 Roma, Italy}
\altaffiltext{10}{Space Sciences Laboratory, 7 Gauss Way, University of California, Berkeley, CA 94720-7450, USA}
\altaffiltext{11}{Columbia Astrophysics Laboratory, 550 West 120th Street, New York, NY 10027, USA}

\begin{abstract} 
We present $\emph{NuSTAR}$ observations of neutron star (NS) low-mass X-ray binaries: 4U 1636-53, GX 17+2, and 4U 1705-44.  We observed 4U 1636-53 in the hard state, with an Eddington fraction, $F_{\mathrm{Edd}}$, of 0.01; GX 17+2 and 4U 1705-44 were in the soft state with fractions of 0.57 and 0.10, respectively. Each spectrum shows evidence for a relativistically broadened Fe K$_{\alpha}$ line. Through accretion disk reflection modeling, we constrain the radius of the inner disk in 4U 1636-53 to be $R_{in}=1.03\pm0.03$  ISCO (innermost stable circular orbit) assuming a dimensionless spin parameter $a_{*}=cJ/GM^{2}=0.0$, and $R_{in}=1.08\pm0.06$ ISCO for $a_{*}=0.3$ (errors quoted at 1 $\sigma$). This value proves to be model independent. For $a_{*}=0.3$ and $M=1.4\ M_{\odot}$, for example, $1.08\pm0.06$ ISCO translates to a physical radius of $R=10.8\pm0.6$ km, and the neutron star would have to be smaller than this radius (other outcomes are possible for allowed spin parameters and masses). For GX 17+2, $R_{in}=1.00-1.04$ ISCO for $a_{*}=0.0$ and $R_{in}=1.03-1.30$ ISCO for $a_{*}=0.3$. For $a_{*}=0.3$ and $M=1.4\ M_{\odot}$, $R_{in}=1.03-1.30$ ISCO translates to $R=10.3-13.0$ km. The inner accretion disk in 4U 1705-44 may be truncated just above the stellar surface, perhaps by a boundary layer or magnetosphere; reflection models give a radius of 1.46-1.64 ISCO for $a_{*}=0.0$ and 1.69-1.93 ISCO for $a_{*}=0.3$. We discuss the implications that our results may have on the equation of state of ultradense, cold matter and our understanding of the innermost accretion flow onto neutron stars with low surface magnetic fields, and systematic errors related to the reflection models and spacetime metric around less idealized neutron stars.
\end{abstract}

\section{Introduction}
The equation of state (EOS) of ultradense matter is not yet known. Earth laboratories are unable to replicate the necessary environment to constrain the EOS. Thus, the only way to study matter under these extreme conditions is through observations of neutron stars (NS). The EOS sets the mass and radius of the NS. Theoretical mass-radius tracks have been compiled for different theoretical models (see \citealt{LP16} for a review). Three-body interactions between nucleons make the radius of the NS the most important quantity in determining the EOS because it does not change quickly as a function of mass. Hence, constraining the radius of the NS can, in turn, lead to a determination of the EOS of the cold, dense matter under extremely dense physical conditions \citep{lattimer11}.

Low-mass X-ray binaries (LMXBs) are one setting in which we can attempt to constrain the EOS through observations. LMXBs consist of a roughly stellar-mass companion that transfers matter onto a compact object via Roche-lobe overflow. 
Broad iron line profiles have been seen in these accreting systems that harbor a black hole (BH; e.g. \citealt{Fabian89}; \citealt{miller02}, \citeyear{miller07}; \citealt{reis08}, \citeyear{reis09a}) or NS  (e.g. \citealt{BS07}; \citealt{papitto08}; \citealt{cackett08}, \citeyear{cackett09}, \citeyear{cackett10}; \citealt{disalvo09}, \citeyear{disalvo15}; \citealt{Egron13}; \citealt{miller13}) as the primary accreting compact object. LMXBs are ideal laboratories for conducting radius measurements through reflection studies since all accretion occurs via the disk, as opposed to their high-mass X-ray binary (HMXB) counterparts which can accrete via stellar winds.

X-ray disk lines are produced from an external hard X-ray source irradiating the accretion disk. In the case of NSs, the hard X-ray source could be a hot corona, the stellar surface, or a boundary layer and may be thermal or non-thermal in nature. Regardless of the nature of the hard X-ray emission, the asymmetrically broadened profile of the Fe K$_{\alpha}$ line gives a direct measure of the position of inner disk since the effects of gravitational redshift and Doppler redshift/boosting on the emission line become stronger closer to the compact object \citep{Fabian89}. 

The Fe K$_{\alpha}$ line in NS LMXBs can set an upper limit for the radius of NS since the disk must truncate at or before the surface (\citealt{cackett08}, \citeyear{cackett10}; \citealt{reis09b}; \citealt{miller13}; \citealt{degenaar15}). Since the magnetic field could also truncate the accretion disk, studies of disk reflection can also be used to set an upper limit on the magnetic field strength (\citealt{cackett09}; \citealt{Pap09}; \citealt{miller11}; \citealt{degenaar14}, \citeyear{degenaar16}; \citealt{king16}; \citealt{ludlam16}).

We analyze $\emph{NuSTAR}$ \citep{nustar} observations of NS LMXBs 4U1636-53, GX 17+2, and 4U1705-44. 4U 1636-53 and 4U 1705-44 are categorized as atoll sources while GX 17+2 is a \lq \lq Z" source,  according to the classification of \citet{hasinger89}. Atoll sources have lower luminosities ($\sim0.01-0.2\ L_{\mathrm{Edd}}$) than Z sources (see \citet{done07} for a review). The Z sources have soft X-ray spectra that can be described by two thermal components (multicolor disk blackbody and single temperature blackbody; \citealt{lin07}). Atolls can have soft or hard spectra with transitional phases in between. The hard state can be modeled well by a powerlaw and thermal component when needed \citep{lin07}. 

Further, the spectral state may be associated with the level of disk truncation \citep{done07}. A study of 4U 1608-52 by \citet{gd02} found that at low luminosity the accretion disk was truncated while at high luminosity the inner radius of the disk moved inwards. \citet{pintore16} found similar behavior for SAX J1748.9-2021. However, \citet{sanna14} found that the inner disk did not seem to be correlated with the spectral state for 4U 1636-53. The inner disk radius of Serpens X-1 (Ser X-1) also does not appear to change much for a range of luminosities \citep{chiang16b}. 

4U 1636-53 was in the hard state at the time of the $\emph{NuSTAR}$ observation, while GX 17+2 and 4U 1705-44 were in the soft state.
We focus on constraining the inner disk radius in these sources and the implications this has on the EOS for ultradense, cold matter. The following sections (\S 1.1, 1.2, 1.3) give a brief introduction about each source. Section 2 provides details on the observations of each source and how the data were reduced. Section 3 presents our analysis and results. Section 4 discusses those results and we conclude in section 5.

\subsection{4U 1636-53}
 4U 1636-53 is a well studied, persistent LMXB that exhibits Type 1 X-ray bursts and has a spin frequency of 581 Hz (\citealt{zhang97}; \citealt{stroh02}). The source is located a maximum of 6.6 kpc away from inspection of the Type 1 X-ray bursts (assuming the brightest type-I X-ray bursts hit the Eddington limit; \citealt{Galloway08}). The companion star is a 0.4 $M_{\odot}$, 18th magnitude star with an orbital period of 3.8 hours (\citealt{perdersen82}; \citealt{van90}). kHz quasi-periodic oscillations (QPOs) suggest that 4U 1636-53 may harbor a NS as large as 2 $M_{\odot}$ (\citealt{barret06}; \citealt{casares06}). The system regularly undergoes state transitions on $\sim40$ day time intervals \citep{shih05}.

Reflection studies suggest this source has a high inclination, but a lack of dips in the X-ray light curve limits the inclination $\sim70^{\circ}$ (\citealt{frank87}; \citealt{casares06}; \citealt{sanna13}). \citet{pandel08} found the inner disk radius was consistent with the innermost stable circular orbit (ISCO) when looking at observations taken with $\emph{XMM-Newton}$ and $\emph{RXTE}$ while in the transitional and soft state. \citet{cackett10} found larger inner disk radii at $\sim8\pm4$ ISCO (assuming $a_{*}=0$) when applying blurred reflection models to the low flux state, but values  consistent with the ISCO were measured in the soft state. Additionally, \citet{sanna13} analyzed the source in the soft and transitional states with two additional observations. They measure the inner disk radius to be as large as 4.45 ISCO in low flux states, which is smaller but consistent with \citet{cackett10}. \citet{Lyu14} used observations from $\emph{Suzaku}$, $\emph{XMM-Newton}$, and $\emph{RXTE}$ to see if the inner disk radius inferred from Fe line changed over flux states. They used available disk line models ({\sc diskline, laor, kryline}) to account for relativistic broadening and found that the Fe line did not change significantly. They conclude that the line is broadened by mechanisms other than just relativistic broadening, though the data are consistent with a disk remaining at the ISCO if timing parameters trace the mass accretion rate rather than the inner radius.

\subsection{GX 17+2}
GX 17+2 is a burster located a maximum distance of 13.0 kpc \citep{Galloway08} with a spin frequency of 293.2 Hz \citep{wij97}. The counterpart of GX 17+2 is not known currently. It may be NP Ser \citep{tar72} or star \lq \lq A" proposed by \citet{deutsch99} from optical and IR variability studies. However, \citet{callanan02} proposed that the IR variability could be explained by synchrotron flares. 
The system has an inclination of less than $45^{\circ}$ (\citealt{cackett10}, \citeyear{cackett12}). \citet{disalvo00} performed the first extensive spectral analysis of GX 17+2. They were able to limit the radius of the NS between 10-19 km based upon comptonization of photons within their spectra. \citet{cackett10} found a similarly small limit through modeling the Fe reflection in the accretion disk.

\subsection{4U 1705-44}
4U 1705-44 is a persistently bright source that shows Type 1 X-ray bursts \citep{langmeier87} which place it at a maximum distance of 7.8 kpc \citep{Galloway08}. The source has been observed in both the hard and soft states (\citealt{barret02}; \citealt{piraino07}). Broad Fe emission has been in each of these states. The inclination of the system is between $20^{\circ}-50^{\circ}$ \citep{piraino07}.  Di Salvo et al.\ (2009) found evidence for multiple emission lines for SXVI, Ar XVIII, and Ca XIX in addition to iron in observations taken with $\emph{XMM-Newton}$. These lines gave an inner disk radius of $2.3\pm0.3$ ISCO (assuming $a_{*}=0$). \citet{reis09b} found an inner disk radius of $1.75_{-0.28}^{+0.17}$ ISCO when using observations obtained with $\emph{Suzaku}$. \citet{cackett10} used both $\emph{XMM-Newton}$ and $\emph{Suzaku}$ observations and found that the inner disk ranged from 1.0-6.5 ISCO. The inner disk of 4U 1705-44 appears to be truncated in many studies though the degree of truncation varies. 

\section{Observations and Data Reduction}
\begin{table}
\caption{$\emph{NuSTAR}$ Observation Information}
\label{tab:edd} 
\begin{center}
\begin{tabular}{lcccc}
\hline
Source & Obs ID &Date & Net Exp (ks) & Net Rate (cts/s)\\
\hline
4U 1636-53 & 30101024002& 06/06/2015 & 19.2 & 40\\
GX 17+2 &30101023002& 03/22/2016 & 23.3 & 361\\
4U 1705-44 &30101025002& 04/22/2016 & 28.7& 174\\
\hline
\end{tabular}
\end{center}
\end{table}

The details of each $\emph{NuSTAR}$ observation are listed in Table 1. Lightcurves and spectra were created using a 120$^{\prime \prime}$ circular extraction region centered around the source using the {\sc{nuproducts}} tool from {\sc nustardas} v1.5.1 with {\sc caldb} 20160421. A background was generated and subtracted using another region of the same dimension in a region away from the source. There were a total of 6 Type 1 X-ray bursts present in the lightcurves for 4U 1636-53 and a single Type 1 X-ray burst for 4U 1705-44. We will report on the bursts in a separate paper. We created good time intervals (GTIs) to remove $\sim$10 -- 150 s after the initial fast rise (depending on the duration of the individual burst) to separate these from the steady emission. These GTIs were applied during the generation of the spectra for both the FPMA and FPMB. Initial modeling of the persistent emission spectra with a constant, fixed to 1 for the FPMA and allowed to float for the FPMB, found the floating constant to be within 0.95-1.05 in each case. We proceeded to combine the two source spectra, background spectra, and ancillary response matrices via {\sc addascaspec}. We use {\sc addrmf} to create a single redistribution matrix file. The spectra were grouped using {\sc grppha} to have a minimum of 25 counts per bin \citep{cash}.

\section{Spectral Analysis and Results}
We use XSPEC version 12.8.1 \citep{arnaud96} in this work. All errors
are quoted at $1\sigma$ (68$\%$) confidence level and were
  calculated from Monte Carlo Markov Chain (MCMC) of length
  100,000. We perform fits over the energy range in which the
spectrum is not background dominated for each source. We account for
the neutral column hydrogen density along the line of sight via
  {\sc tbabs} \citep{wilms00}. The solar abundance was set to {\sc
  wilm} \citep{wilms00} and {\sc vern} cross sections \citep{Vern96}
were used.

 We experimented with different phenomenological one, two, and
three-component models to describe the spectral continua.  None of the
sources required three components.  The continuum of the spectrum
obtained from 4U 1636$-$53 was well described with a cut-off power-law
model.  The cut-off energy may reflect the electron temperature of the
corona, and the simple continuum component may only be an
approximation of a region dominated by Comptonization.  The spectral
continua of GX 17$+$2 and 4U 1705$-$44 were well described by a
combination of disk blackbody \citep{diskbb} and simple
blackbody components.  It is possible to infer radii from both disk
blackbody and simple blackbody components; however, it is also clear
that scattering can harden emergent spectra and cause falsely small
radii to be inferred (e.g., \citealt{london}; \citealt{shimura}; \citealt{merloni00}).  Low temperature, optically-thick
Comptonization can sometimes be approximated as a blackbody; it is
likely that our simple blackbody component accounts for this process
in the boundary layer (e.g., \citealt{gilfanov03}).  Given the many
complexities and possible distortions, we do not look to the continua
for robust physical inferences; rather, we utilize disk reflection for
this purpose.

To properly describe the reflection features and relativistic effects
present in the data for 4U 1636-53 we employ the model {\sc relxill}
\citep{garcia14}. This model self-consistently accounts for X-ray
reflection and relativistic ray tracing for a power law irradiating an
accretion disk while properly taking into account the inclination in
the reflection from the disk. The parameters of this model are as
follows: inner emissivity ($q_{in}$), outer emissivity ($q_{out}$),
break radius ($R_{break}$) between the two emissivities, spin
parameter ($a_{*}=cJ/GM^{2}$), inclination if the disk ($i$), inner
radius of the disk ($R_{in}$) in units of ISCO, outer radius of the
disk ($R_{out}$), redshift ($z$), photon index of the power
law($\Gamma$), log of the ionization parameter ($log(\xi)$), iron
abundance ($A_{Fe}$), cut off energy of the power law ($E_{cut}$),
reflection fraction ($f_{refl}$), and normalization.

We apply {\sc bbrefl} \citep{bbrefl} to GX 17+2 and 4U 1705-44 in order to model the emergent reflection emission from a disk assuming an irradiating blackbody continuum, and convolve it with {\sc relconv} \citep{relconv}. The iron abundance in {\sc bbrefl} comes in two flavors: solar abundance and twice solar abundance. The parameters of {\sc bbrefl} are as follows: log of the ionization parameter ($log(\xi)$), temperature of the incident blackbody in keV ($kT$), iron abundance ($A_{Fe}$), reflection fraction ($f_{refl}$), redshift ($z$), and normalization. The parameters of {\sc relconv} are as follows: inner emissivity index ($q_{in}$),  outer emissivity index ($q_{out}$), dimensionless spin parameter ($a_{*}$), inner disk radius in units of ISCO ($R_{in}$), and outer disk radius in units of gravitational radii ($R_{out}$). 

 Additionally, we use {\sc reflionx} \citep{reflionx} to test the robustness of our results obtained with {\sc relxill} and the nature of the Fe K$_{\alpha}$ line in our {\sc bbrefl}. In the case of the 4U 1636-53, we use a version of {\sc reflionx}\footnote{https://www-xray.ast.cam.ac.uk/$\sim$mlparker/reflionx\_models/reflionx\_hc.mod} that has a variable high energy cut-off. We convolve this model with the relativistic blurring kernel {\sc kerrconv} \citep{kerrconv}. This provides a completely model independent check to values obtained with {\sc relxill}. In the case of GX 17+2 and 4U 1705-44, we have a version of {\sc reflionx}\footnote{http://www-xray.ast.cam.ac.uk/$\sim$mlparker/reflionx$\_$models/reflionx$\_$bb.mod} that has been modified to assume a blackbody is illuminating the disk. This reflection model does not represent an independent check on {\sc bbrefl} since they are derived from the same parent code \citep{ross93}. It can, however, be used to verify that the Fe K$_{\alpha}$ line is dynamically broadened and not a result of gas broadening. This is because the {\sc reflionx} model contains additional physics that accounts for gas effects. It has a broader range of elements, charge states, and ionization while it accounts for the local radiation field at each point. Hence, if we are able to obtain similar values for various parameters, then the line profile is dynamical in origin.
 
A few reasonable conditions were enforced when making fits with {\sc relxill} and {\sc relconv}. First, we tie the outer emissivity index, $q_{out}$ to the inner emissivity index, $q_{in}$, to create a constant emissivity index. Next, we fix the spin parameter, $a_{*}$ (where $a_{*}=cJ/GM^{2}$), in the model {\sc relconv} to 0.0 and 0.3 in the subsequent fits since NS in LMXBs have $a_{*} \leq 0.3$ (\citealt{miller11}; \citealt{Galloway08}). This does not hinder our estimate of the inner radius since the position of the ISCO is relatively constant for low spin parameters (corrections for frame-dragging for $a_{*}<0.3$ give errors $\ll10\%$; \citealt{Miller98}). Further, the outer disk radius has been fixed to 990 $R_{g}$ (where  $R_{g} = GM/c^{2}$). In the case that we use {\sc kerrconv}, we tie the emissivity indices to create one emissivity index and fix the outer disk radius to 400 ISCO.

\begin{table}
\caption{4U 1636 Relxill Modeling for Different Spin Parameters}
\label{tab:relxill} 
\begin{center}
\begin{tabular}{llcc}
\hline
Component & Parameter & \multicolumn{2}{c}{\sc relxill} \\
\hline
{\sc tbabs}
&$N_\mathit{H} (10^{22}) ^{\dagger}$
&$0.3$
&$0.3$
\\
{\sc relxill}
&$q$
&$2.25\pm0.05$
&$2.19\pm0.04$
\\
&$a_{*} ^{\dagger}$
&$0.0$
&$0.3$
\\
&$\mathit{i} (^{\circ})$
&$78.2\pm1.67$
&$78.5\pm1.22$
\\
&$R_\mathit{in} (ISCO) $
&$1.03\pm0.03$
&$1.08\pm0.06$
\\
&$R_\mathit{out} (R_\mathit{g}) ^{\dagger}$ 
&990
&990
\\
&$R_\mathit{in} (km) $
&$12.4\pm0.4$
&$10.8\pm0.6$
\\
&$\mathit{z} ^{\dagger}$
&0.0
&0.0
\\
&$\Gamma$
&$1.74\pm0.01$
&$1.74\pm0.01$
\\
&$log(\xi)$
&$3.3\pm0.1$
&$3.26\pm0.06$
\\
&$A_\mathit{Fe}$
&$4.9\pm0.1$
&$4.8\pm0.1$
\\
&$E_\mathit{cut} (keV)$
&$20.5\pm0.3$
&$20.5\pm0.3$
\\
&$\mathit{f}_\mathit{refl}$
&$0.42\pm0.05$
&$0.40\pm0.04$
\\
&norm $(10^{-3})$
&$2.21\pm0.06$
&$2.22\pm0.03$
\\
&$F_{\mathrm{unabs},3.0-50.0\ keV}$
&$8.9\pm0.2$
&$8.9\pm0.1$
\\
&$L_{\mathrm{unabs},3.0-50.0\ keV}$
&$4.6\pm0.1$
&$4.63\pm0.06$
\\
&$F_{\mathrm{unabs},0.5-50.0\ keV}$
&$12.8\pm0.3$
&$12.8\pm0.2$
\\
&$L_{\mathrm{unabs},0.5-50.0\ keV}$
&$6.7\pm0.2$
&$6.67\pm0.09$
\\
\hline
&$\chi_\nu^{2}$(dof)
&1.06 (862) 
&1.06 (862)  
\\
\hline
$^{\dagger}$ = fixed
\end{tabular}

\medskip
Note.--- Errors are quoted at $1\sigma$ confidence level. The absorption column density was fixed to the \citet{dl90} value and given in units of cm$^{-2}$. The spin parameter is pegged at an upper limit of 0.3. The inner disk radius in units of km assumes a NS mass of 1.4 $M_{\odot}$. Flux is given in units of $10^{-10}$ ergs cm$^{-2}$ s$^{-1}$. Luminosity is calculated at a maximum of 6.6 kpc and given in units of $10^{36}$ ergs s$^{-1}$.
\end{center}
\end{table}

\begin{table}
\caption{4U 1636 Reflionx Modeling for Different Spin Parameters}
\label{tab:relxill} 
\begin{center}
\begin{tabular}{llcc}
\hline
Component & Parameter & \multicolumn{2}{c}{\sc reflionx} \\
\hline
{\sc tbabs}
&$N_\mathit{H} (10^{22}) ^{\dagger}$
&$0.3$
&$0.3$
\\
{\sc cutoffpl}
&$\Gamma$
&$1.74\pm0.01$
&$1.75\pm0.01$
\\
&$E_\mathit{cut} (keV)$
&$20.7\pm0.3$
&$20.6\pm0.3$
\\
&norm 
&$0.20\pm0.01$
&$0.20\pm0.01$
\\
{\sc kerrconv}
&$q$
&$2.33\pm0.04$
&$2.28\pm0.05$
\\
&$a_{*} ^{\dagger}$
&0.0
&0.3
\\
&$\mathit{i} (^{\circ})$
&$78.6\pm1.2$
&$78.3\pm1.2$
\\
&$R_\mathit{in} (ISCO) $
&$1.02\pm0.02$
&$1.08\pm0.06$
\\
&$R_\mathit{out} (ISCO) ^{\dagger}$ 
&400
&400
\\
&$R_\mathit{in} (km) $
&$12.2\pm0.2$
&$10.8\pm0.6$
\\
{\sc reflionx}
&$\xi$
&$1800\pm600$
&$1100\pm500$
\\
&$A_\mathit{Fe} $
&$4.4\pm0.5$
&$4.6\pm0.3$
\\
&$f_{refl}$
&$0.6\pm0.2$
&$0.6\pm0.3$
\\
&$\mathit{z} ^{\dagger}$
&0
&0
\\
&norm $(10^{-7})$
&$2.7\pm0.9$
&$6.8\pm4.0$
\\
&$F_{\mathrm{unabs, 3.0-50.0\ keV}}$
&$9\pm3$
&$9\pm5$
\\
&$L_{\mathrm{unabs, 3.0-50.0\ keV}}$
&$5\pm2$
&$5\pm3$
\\
&$F_{\mathrm{unabs, 0.5-50.0\ keV}}$
&$14\pm5$
&$14\pm8$
\\
&$L_{\mathrm{unabs, 0.5-50.0\ keV}}$
&$7\pm2$
&$7\pm4$
\\
\hline
&$\chi_\nu^{2}$(dof)
&1.25 (862) 
&1.23 (862)
\\
\hline
$^{\dagger}$ = fixed
\end{tabular}

\medskip
Note.--- Errors are quoted at $1\sigma$ confidence level. The absorption column density was fixed to the \citet{dl90} value and given in units of cm$^{-2}$.  The spin parameter is pegged at an upper limit of 0.3. The {\sc reflionx} model used has a variable high energy cut off which we tied to the cut off energy of the cut off power law used to model the continuum emission. Additionally, we tied the photon index between {\sc reflionx} and {\sc cutoffpl}. The inner disk radius in units of km assumes a NS mass of 1.4 $M_{\odot}$. Flux is given in units of $10^{-10}$ ergs cm$^{-2}$ s$^{-1}$. Luminosity is calculated at a maximum of 6.6 kpc and given in units of $10^{36}$ ergs s$^{-1}$.
\end{center}
\end{table}

\begin{figure}
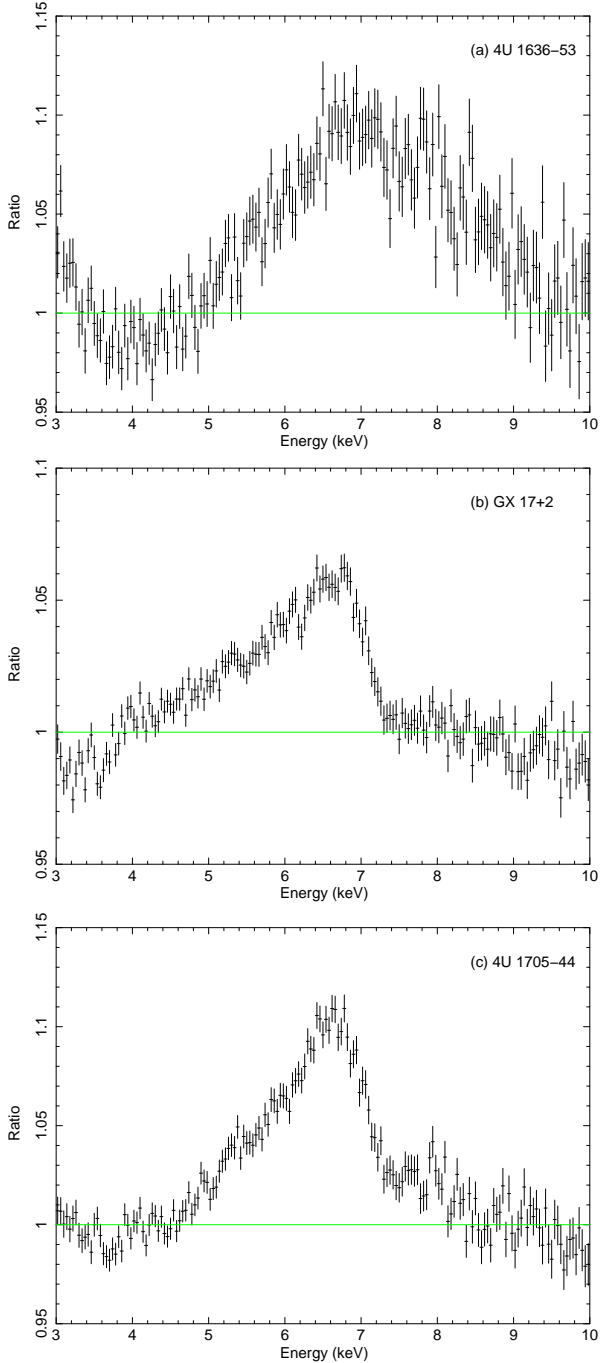

\centering
\includegraphics[angle=270,width=8.4cm]{1636fe3.eps}
\includegraphics[angle=270,width=8.4cm]{17p2fe3.eps}
\includegraphics[angle=270,width=8.4cm]{1705fe3.eps}
\caption{Ratio of the data to the continuum model  in the Fe K band for $\emph{NuSTAR}$ observation of 4U 1636-53, GX 17+2, and 4U 1705-44.  The iron line region from 5-8 keV was ignored to prevent the feature from skewing the fit. The data were rebinned for plotting purposes. (a) A simple cut-off power law was fit over the energies of 3.0-5.0 keV and 8.0-50.0 keV. 
For panels (b) and (c), a simple disk blackbody and single temperature blackbody was fit over the energies of 3.0-5.0 keV and 8.0-30.0 keV.
}
\label{fig:feline}
\end{figure}

\begin{figure}
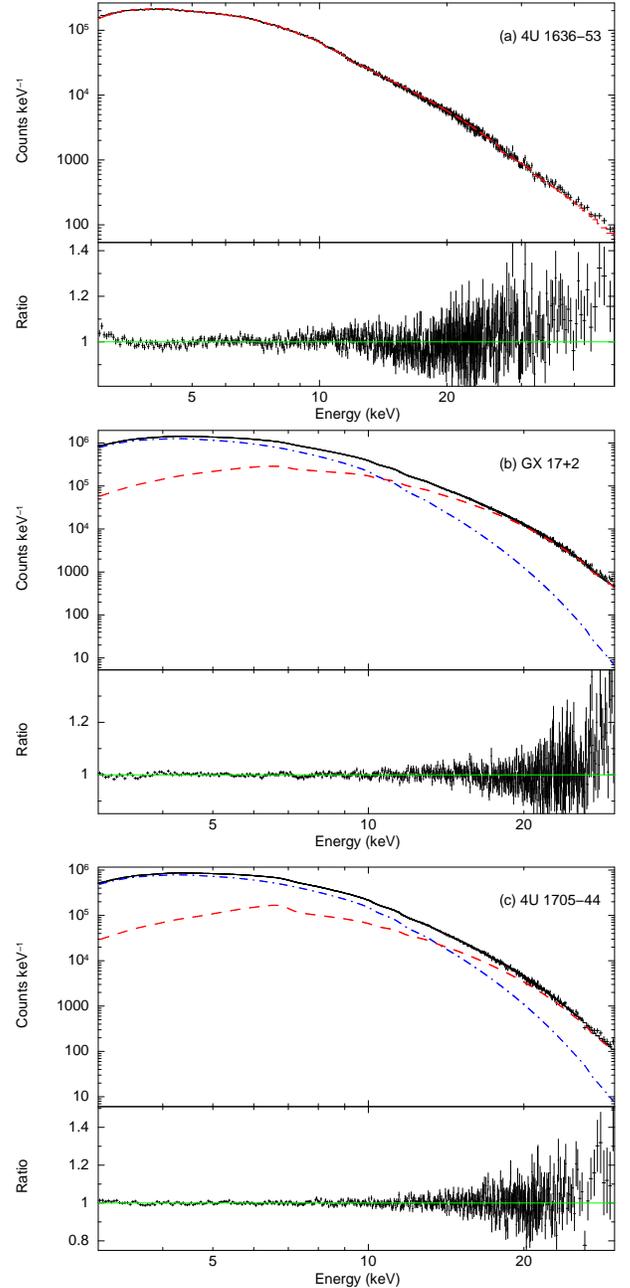

\centering
\includegraphics[angle=270,width=8.4cm]{4u1636spec.eps}
\includegraphics[angle=270,width=8.4cm]{gx17spec.eps}
\includegraphics[angle=270,width=8.4cm]{4u1705spec.eps}
\caption{(a) 4U 1636-53 spectrum fit from 3.0-50.0 keV with {\sc relxill} (red dash line) for the fit in Table 2. The panel below shows the ratio of the data to the model.
(b) GX 17+2 spectrum fit from 3.0-30.0 keV with {\sc diskbb} (blue dot-dash line) and {\sc bbrefl} (red dash line) for the fit in Table 4. The panel below shows the ratio of the data to the model.
(c) 4U 1705-44 spectrum fit from 3.0-30.0 keV with {\sc diskbb} (blue dot-dash line) and {\sc bbrefl} (red dash line) for the fit in Table 5. The panel below shows the ratio of the data to the model.
The data were rebinned for plotting purposes.
}
\label{fig:feline}
\end{figure}

\subsection{4U 1636-53}
As noted above, initial fits were performed from 3.0-50.0 keV with an
absorbed cut-off power law to model the continuum emission.  Although
the continuum is well described, this model gives a poor fit
($\chi^{2}/dof=2435.47/876$) since it does not account for the
reflection present within the spectrum.  A broad Fe K$_{\alpha}$ line
centered $\sim 7$ keV with a red wing extending to lower energies and
blue wing extending up to 9 keV, implying a high inclination, can be
seen in the top panel of Figure 1.  The addition of a thermal
component improved the overall fit, but the temperature was unfeasibly
low for the sensitivity of $\emph{NuSTAR}$ ($0.18\pm0.02$ keV) and the
normalization was not physically possible
($3.6_{-3.4}^{+23.2}\times10^{7}$), hence we excluded it from
subsequent modeling.

We performed two fits with {\sc tbabs}*{\sc relxill} for the different
spin values. Each of these provide a significantly better fit
(28$\sigma$ improvement) over the absorbed cut-off power law and can
be seen in Table 2. The ionization parameter is reasonable for
accreting sources across all fits. The iron abundance is large,
  but fixing it to lower values does not change the inner disk
  radius. Regardless of the iron abundance, each fit consistently
  gives a small inner disk radius of $R_{in}=1.03\pm0.03$ ISCO for
  $a_{*}=0.0$ and $R_{in}=1.08\pm0.06$ ISCO for $a_{*}=0.3$. The fit
  returns a high inclination of $76.5^{\circ}-79.9^{\circ}$ for each
  spin value. The photon index is $1.74\pm0.01$ with a cut off energy
at $\sim20.5$ keV. The best fit spectrum can be seen in Figure
2. Figure 3 shows the goodness-of-fit versus the inner disk radius
(top panel).

To check that the values obtained with our {\sc relxill} modeling of
4U 1636-53 are not model dependent, we apply the model {\sc reflionx}
to characterize the emergent reflection spectrum arising from an
incident cut-off power law continuum. We blur the emergent reflection
with the relativistic convolution model {\sc kerrconv}. This version
of {\sc reflionx} has an adjustable cut-off energy which we tie with
the cut-off power law model used to model the continuum emission. This
model, however, is angle averaged, unlike {\sc relxill}, which
properly takes into account the inclination of the disk when tracing
each photon from the disk.

The resulting fit can be seen in Table 3 for  a spin of 0.0 and
  0.3. The inner disk radius is consistent with the values found in
the {\sc relxill} modeling. Additionally, the photon index, high
energy cut-off, emissivity index, spin parameter, ionization, and
inclination are within error for the values found with {\sc
  relxill}. This demonstrates that the inner disk radius measurement
is robust.

\subsection{GX 17+2}
\begin{table*}
\caption{GX 17+2 Reflection Modeling}
\label{tab:GX} 
\begin{center}
\begin{tabular}{llcccccc}
\hline
Component & Parameter&\multicolumn{4}{c}{\sc bbrefl}&\multicolumn{2}{c}{\sc reflionx}\\
\hline
{\sc tbabs}
&$N_\mathit{H} (10^{22}) ^{\dagger}$
&$0.9$
&$0.9$
&$0.9$
&$0.9$
&$0.9$
&$0.9$
\\
{\sc diskbb}
&$T_{in}$
&$1.93\pm0.04$
&$1.992\pm0.001$
&$1.93\pm0.05$
&$1.92\pm0.06$
&$1.92\pm0.01$
&$1.92\pm0.01$
\\
&$norm$
&$26\pm1$
&$23.8\pm0.1$
&$26\pm2$
&$26\pm2$
&$26.8\pm0.4$
&$26.7\pm0.4$
\\
{\sc blackbody}
&$kT$
&...
&...
&...
&...
&$2.86\pm0.01$
&$2.86\pm0.01$
\\
&norm $(10^{-2})$
&...
&...
&...
&...
&$2.9\pm0.1$
&$2.9\pm0.1$
\\
{\sc relconv}
&$q$
&$8.1_{-1.0}^{+0.8}$
&$3.7_{-0.3}^{+4.7}$
&$6.0\pm3.0$
&$4.5\pm1.5$
&$3.5\pm0.1$
&$3.2\pm0.1$
\\
&$a_{*} ^{\dagger}$
&$0.0$
&$0.3$
&$0.0$
&$0.3$
&$0.0$
&$0.3$
\\
&$\mathit{i} (^{\circ})$
&$35.1_{-0.6}^{+0.2}$
&$30_{-2}^{+8}$
&$30\pm5$
&$30\pm4$
&$25.9\pm0.2$
&$25.8\pm0.2$
\\
&$R_\mathit{in} (ISCO) $
&$1.00^{+0.02}$
&$1.15_{-0.08}^{+0.15}$
&$1.02\pm0.02$
&$1.10\pm0.07$
&$1.01\pm0.01$
&$1.01\pm0.01$
\\
&$R_\mathit{out} (R_\mathit{g}) ^{\dagger}$ 
&990
&990
&990
&990
&990
&990
\\
&$R_\mathit{in} (km) $
&$12.0^{+0.2}$
&$11.5_{-0.8}^{+1.5}$
&$12.2\pm0.2$
&$11.0\pm0.7$
&$12.1\pm0.1$
&$10.1\pm0.1$
\\
{\sc bbrefl}
&$log(\xi)$
&$2.47_{-0.01}^{+0.07}$
&$2.45\pm0.01$
&$2.33_{-0.01}^{+0.09}$
&$2.40\pm0.06$
&...
&...
\\
&$kT$ (keV)
&$3.15_{-0.07}^{+0.01}$
&$3.12_{-0.01}^{+0.04}$
&$3.03_{-0.01}^{+0.05}$
&$3.03\pm0.02$
&...
&...
\\
&$A_\mathit{Fe} ^{\dagger}$
&$1.0$
&$1.0$
&$2.0$
&$2.0$
&...
&...
\\
&$\mathit{f}_\mathit{refl}$
&$1.69_{-0.67}^{+0.04}$
&$1.52_{-0.01}^{+0.67}$
&$1.2_{-0.4}^{+0.2}$
&$1.27_{-0.50}^{+0.04}$
&...
&...
\\
&$z ^{\dagger}$
&0
&0
&0
&0
&...
&...
\\
&norm $(10^{-26})$
&$8.83_{-0.01}^{+0.60}$
&$8.61_{-0.36}^{+0.02}$
&$13.0^{+1.0}_{-0.03}$
&$13.1_{-0.01}^{+0.08}$
&...
&...
\\
{\sc reflionx}
&$\xi$
&...
&...
&...
&...
&$240\pm20$
&$230\pm7$
\\
&$A_\mathit{Fe}$
&...
&...
&...
&...
&$2.1\pm0.4$
&$2.1\pm0.4$
\\
&$f_{refl}$
&...
&...
&...
&...
&$0.9\pm0.1$
&$0.9\pm0.1$
\\
&$\mathit{z} ^{\dagger}$
&...
&...
&...
&...
&0
&0
\\
&norm
&...
&...
&...
&...
&$1.7\pm0.2$
&$1.8\pm0.1$
\\
&$F_{\mathrm{unabs, 3.0-30.0 keV}}$
&$7.3^{+0.6}_{-0.3}$
&$7.30_{-0.31}^{+0.04}$
&$7.3_{-0.6}^{+0.8}$
&$7.3\pm0.1$
&$7.3\pm0.9$
&$7.3\pm0.5$
\\
&$L_{\mathrm{unabs, 3.0-30.0 keV}}$
&$1.5\pm0.1$
&$1.48_{-0.06}^{+0.01}$
&$1.5_{-0.1}^{+0.2}$
&$1.48\pm0.02$
&$1.5\pm0.2$
&$1.5\pm0.1$
\\
&$F_{\mathrm{unabs, 0.5-30.0 keV}}$
&$10.7_{-0.4}^{+0.8}$
&$10.7_{-0.4}^{+0.1}$
&$10.7_{-0.8}^{+1.2}$
&$10.7\pm0.8$
&$11.0\pm1.3$
&$11.0\pm0.7$
\\
&$L_{\mathrm{unabs, 0.5-30.0 keV}}$
&$2.2_{-0.1}^{+0.2}$
&$2.16_{-0.08}^{+0.01}$
&$2.2\pm0.2$
&$2.2\pm0.2$
&$2.2\pm0.3$
&$2.2\pm0.1$
\\
\hline
&$\chi_\nu^{2}$(dof)
&1.06 (665)
&1.13 (665)
&1.19 (665)
&1.18 (665)
&1.28 (664) 
&1.26 (664) 

\\
\hline
$^{\dagger}$ = fixed
\end{tabular}

\medskip
Note.--- Errors are quoted at $1\sigma$ confidence level. The absorption column density was fixed to the \citet{dl90} value and given in units of cm$^{-2}$. Limb brightening was assumed. The emissivity index was pegged at the hard limit of 3.0. The {\sc reflionx} model used has been modified for a blackbody illuminating the accretion disk. The inclination for {\sc reflionx} was pegged within the limits found with the {\sc bbrefl} due to the model being angle averaged and not able to constrain the inclination. The ionization parameter was also pegged within the {\sc bbrefl} values since it was unconstrained on its own. The blackbody temperature in {\sc reflionx} was tied to the temperature of the blackbody used to model the continuum emission. The emissivity index was pegged at the hard limit of 3.0. The inner disk radius in units of km assumes a NS mass of 1.4 $M_{\odot}$. Flux is given in units of $10^{-9}$ ergs cm$^{-2}$ s$^{-1}$. Luminosity is calculated at a maximum of 13.0 kpc and given in units of $10^{38}$ ergs s$^{-1}$.
\end{center}
\end{table*}

Initial fits were performed from 3.0-30.0 keV with an absorbed single
temperature blackbody and a multi-temperature blackbody component.  This
gives a poor fit ($\chi^{2}/dof=4289.59/670$) because the reflection
spectrum is not yet modeled. The Fe K$_{\alpha}$ emission can be seen
in the middle panel in Figure 1. The red wing extends down to $\sim 4$
keV while the blue wing drops around $\sim7$ keV.

We use {\sc bbrefl} to model the emergent reflection emission and
convolve it with {\sc relconv}. The overall model we used was {\sc
  tbabs}*({\sc diskbb}+{\sc relconv}*{\sc bbrefl}). This model
  provides a better fit with $\chi^{2}/dof=793.2/665$ (33$\sigma$
  improvement for the highest $\chi_{\nu}^{2}$). Parameters and values
  can be seen in Table 4. Figure 2 shows the best fit spectrum. For
  $a_{*}=0.0$, the inner disk radius is tightly constrained to
  $R_{in}=1.00-1.02$ ISCO for $A_{Fe}=1.0$ and $R_{in}=1.00-1.04$ ISCO
  for $A_{Fe}=2.0$. For the higher value of spin, $a_{*}=0.3$,
  $R_{in}=1.15_{-0.08}^{+0.15}$ ISCO for $A_{Fe}=1.0$ and
  $R_{in}=1.10\pm0.07$ ISCO for $A_{Fe}=2.0$. The emissivity index is high
  in each case (ray tracing assuming either a hot spot at a modest
  latitude, or a heated equatorial region, both predict $q=3.0$;
  D. Wilkins, priv. comm.). The inclination lies between
  $25^{\circ}-38^{\circ}$ for all fits. Figure 3 shows the change in
goodness-of-fit versus the inner disk radius (middle panel).

To test the origin of the Fe K$_{\alpha}$ line, we use a version of
{\sc reflionx} that assumes a blackbody is illuminating the accretion
disk. Table 4 shows parameters and values using this model. The
  overall fit is worse, but still requires a small inner disk radius of
  $R_{in}=1.00-1.02$ ISCO for both values of spin. The inclination and
  ionization is consistent with the values found in the previous fits
  with {\sc bbrefl}. The emissivity index is close to simple
  expectations. Regardless, the small inner radii measured with {\sc
  reflionx} imply that dynamical broadening is dominant in shaping the
iron line, and any atmospheric effects are secondary.

\begin{table*}
\caption{4U 1705-44 Reflection Modeling}
\label{tab:1705} 
\begin{center}
\begin{tabular}{llcccccc}
\hline
Component & Parameter&\multicolumn{4}{c}{\sc bbrefl}&\multicolumn{2}{c}{\sc reflionx}\\
\hline
{\sc tbabs}
&$N_\mathit{H} (10^{22}) ^{\dagger}$
&$0.7$
&$0.7$
&$0.7$
&$0.7$
&$0.7$
&$0.7$
\\
{\sc diskbb}
&$T_{in}$
&$2.13\pm0.01$
&$2.13\pm0.01$
&$2.00\pm0.01$
&$2.00\pm0.01$
&$1.95\pm0.02$
&$1.95\pm0.03$
\\
&$norm$
&$9.65\pm0.02$
&$9.6\pm0.1$
&$11.48\pm0.02$
&$11.60\pm0.01$
&$12.3\pm0.5$
&$12.4\pm0.6$
\\
{\sc blackbody}
&$kT$
&...
&...
&...
&...
&$2.54\pm0.02$
&$2.54\pm0.03$
\\
&norm $(10^{-2})$
&...
&...
&...
&...
&$1.15\pm0.05$
&$1.15\pm0.08$
\\
{\sc relconv}
&$q$
&$3.1\pm0.1$
&$3.1\pm0.1$
&$3.2\pm0.1$
&$3.2\pm0.1$
&$2.8\pm0.3$
&$2.7\pm0.3$
\\
&$a_{*} ^{\dagger}$
&$0.0$
&$0.3$
&$0.0$
&$0.3$
&$0.0$
&$0.3$
\\
&$\mathit{i} (^{\circ})$
&$24.4\pm0.4$
&$24.4\pm0.4$
&$24.6\pm0.5$
&$25.6\pm0.5$
&$13\pm11$
&$16\pm6$
\\
&$R_\mathit{in} (ISCO) $
&$1.54\pm0.08$
&$1.82\pm0.09$
&$1.50\pm0.07$
&$1.78\pm0.09$
&$1.21\pm0.18$
&$1.37\pm0.27$
\\
&$R_\mathit{out} (R_\mathit{g}) ^{\dagger}$ 
&990
&990
&990
&990
&990
&990
\\
&$R_\mathit{in} (km) $
&$18.4\pm1.0$
&$18.2\pm0.9$
&$18.0\pm0.8$
&$17.8\pm0.9$
&$14.5\pm2.2$
&$13.7\pm2.7$
\\
{\sc bbrefl}
&$log(\xi)$
&$2.67\pm0.02$
&$2.66\pm0.02$
&$2.74\pm0.04$
&$2.74\pm0.04$
&...
&...
\\
&$kT$ (keV)
&$2.83\pm0.01$
&$2.84\pm0.02$
&$2.67\pm0.01$
&$2.67\pm0.01$
&...
&...
\\
&$A_\mathit{Fe} ^{\dagger}$
&$1.0$
&$1.0$
&$2.0$
&$2.0$
&...
&...
\\
&$\mathit{f}_\mathit{refl}$
&$2.0\pm0.1$
&$2.1\pm0.2$
&$0.74\pm0.02$
&$0.74\pm0.02$
&...
&...
\\
&$z ^{\dagger}$
&0
&0
&0
&0
&...
&...
\\
&norm $(10^{-26})$
&$1.24\pm0.06$
&$1.24\pm0.08$
&$2.1\pm0.2$
&$2.1\pm0.2$
&...
&...
\\
{\sc reflionx}
&$\xi$
&...
&...
&...
&...
&$430\pm20$
&$430\pm20$
\\
&$A_\mathit{Fe}$
&...
&...
&...
&...
&$2.0\pm0.4$
&$2.0\pm0.6$
\\
&$f_{refl}$
&...
&...
&...
&...
&$0.72\pm0.06$
&$0.72\pm0.08$
\\
&$\mathit{z} ^{\dagger}$
&...
&...
&...
&...
&0
&0
\\
&norm
&...
&...
&...
&...
&$0.57\pm0.04$
&$0.60\pm0.05$
\\
&$F_{\mathrm{unabs, 3.0-30.0 keV}}$
&$3.4\pm0.2$
&$3.4\pm0.2$
&$3.4\pm0.3$
&$3.4\pm0.3$
&$3.4\pm0.3$
&$3.4\pm0.4$
\\
&$L_{\mathrm{unabs, 3.0-30.0 keV}}$
&$2.5\pm0.1$
&$2.5\pm0.2$
&$2.5\pm0.2$
&$2.5\pm0.2$
&$2.5\pm0.2$
&$2.5\pm0.3$
\\
&$F_{\mathrm{unabs, 0.5-30.0 keV}}$
&$5.0\pm0.2$
&$5.0\pm0.3$
&$5.0\pm0.5$
&$5.0\pm0.5$
&$5.2\pm0.5$
&$5.2\pm0.6$
\\
&$L_{\mathrm{unabs, 0.5-30.0 keV}}$
&$3.7\pm0.2$
&$3.7\pm0.2$
&$3.7\pm0.3$
&$3.7\pm0.3$
&$3.8\pm0.3$
&$3.8\pm0.5$
\\
\hline
&$\chi_\nu^{2}$(dof)
&1.16 (665)
&1.16 (665)
&1.19 (665)
&1.19 (665)
&1.21 (664) 
&1.21 (664) 

\\
\hline
$^{\dagger}$ = fixed
\end{tabular}

\medskip
Note.--- Errors are quoted at $1\sigma$ confidence level. The absorption column density was fixed to the \citet{dl90} value and given in units of cm$^{-2}$. Limb brightening was assumed. The {\sc reflionx} model used has been modified for a blackbody illuminating the accretion disk. The inclination for {\sc reflionx} was pegged within the limits found with the {\sc bbrefl} due to the model being angle averaged and not able to constrain the inclination. The ionization parameter was also pegged within the {\sc bbrefl} values since it was unconstrained on its own. The blackbody temperature in {\sc reflionx} was tied to the temperature of the blackbody used to model the continuum emission. Inner disk radius in units of km assumes a NS mass of 1.4 $M_{\odot}$. 
Flux is given in units of $10^{-9}$ ergs cm$^{-2}$ s$^{-1}$. Luminosity is calculated at a maximum of 7.8 kpc and given in units of $10^{37}$ ergs s$^{-1}$.
\end{center}
\end{table*}

\subsection{4U 1705-44}
Initial fits were performed from 3.0-30.0 keV with an absorbed single
temperature blackbody and a
multi-temperature blackbody component. This gives a poor fit ($\chi^{2}/dof=4504.9/670$) since the
strong disk reflection spectrum has not been modeled. See the bottom
panel of Figure 1 for the Fe K$_{\alpha}$ line.

We use {\sc bbrefl} to model the emergent reflection emission and
convolve it with {\sc relconv} \citep{relconv}. The overall model we
used was {\sc tbabs}*({\sc diskbb}+{\sc relconv}*{\sc bbrefl}). 
  This provides a better overall fit (34$\sigma$ improvement for the
  highest $\chi_{\nu}^{2}$). Parameters and values can be seen in
  Table 5. The inner disk radius is slightly truncated prior to the
  neutron star surface between 1.46-1.64 ISCO for $a_{*}=0.0$ for both
  values of iron abundance. For $a_{*}=0.3$, the inner disk radius is
  truncated at 1.69-1.93 ISCO. The inclination is between
  $24.0^{\circ}-26.1^{\circ}$ for all fits. The change in
goodness-of-fit versus the inner disk radius can be seen in the bottom
panel of Figure 3.

Again, we test the origin of our Fe K$_{\alpha}$ line in this
truncated disk with the version {\sc reflionx} that has been modified
for a blackbody illuminating the disk. The resulting fits can be seen
in Table 5.  Even though the overall fit is slightly worse, the
  disk still requires a truncation of $R_{in}=1.21\pm0.18$ ISCO for
  $a_{*}=0.0$ and $R_{in}=1.37\pm0.27$ ISCO for $a_{*}=0.3$. Neither
  model strongly requires that the disk extends to the ISCO, though
  the ISCO is within 2 sigma of the nominal best-fit values found with
  {\sc reflionx}. This confirms that dynamical broadening is the
dominant mechanism for the iron line shape.

 Since the disk does not extend down to the ISCO, we calculate the
  extent of a possible boundary layer and place an upper limit on the
  magnetic field strength since these are plausible scenarios for disk
  truncation.  \citet{PS01} lay out the Newtonian framework for
  boundary layer behavior for different mass accretion rates. We
  estimate the mass accretion for 4U 1705-44 to be
  $(3.4\pm0.4)\times10^{-9}$ M$_{\odot}$ yr$^{-1}$ from the 0.5-30.0
  keV unabsorbed luminosity and using an efficiency of $\eta=0.2$
    \citep{SS00}. Using Equation (25) in \citet{PS01}, we estimate
  that the boundary layer extends out to $\sim1.2$ ISCO (assuming 1.4
  M$_{\odot}$ \& $a_{*}=0.0$). Additional factors, such as spin and
  viscosity in the layer, can extend this region to be consistent with
  the truncation of the inner disk.

 If the disk is impeded by the magnetosphere, we can place an
upper limit on the strength of the field using the upper limit of
$R_{in}=9.8$ R$_{g}$. Assuming a mass of 1.4 M$_{\odot}$, taking the
distance to be 7.8 kpc, and using the unabsorbed flux from 0.5-30.0
keV of $5.2\times10^{-9}$ erg cm$^{-2}$ s$^{-1}$ as the bolometric
flux, we can determine the magnetic dipole moment, $\mu$, from
Equation (1) taken from \citet{cackett09}.   If we make the same
  assumptions about geometry (i.e. $k_{A}=1$, $f_{ang}=1$) and use an
  accretion efficiency of $\eta=0.2$, then
$\mu\simeq2.2\times10^{26}$ G cm$^{3}$. This corresponds to a magnetic
field strength of $B\simeq4.3\times10^{8}$ G at the magnetic poles for
a NS of 10 km. Moreover, if we assume a different conversion factor
$k_{A}=0.5$ \citep{long05} then the magnetic field strength at the
poles would be $B\simeq1.5\times10^{9}$ G. Note that the magnetic
field strength at the pole is twice as strong as at the
equator. However, the type 1 X-ray burst that occurred during the
observation mean material is still reaching the surface of the NS
  and no pulsations have been seen.

\begin{figure}[h!]
\centering
\includegraphics[width=8.4cm]{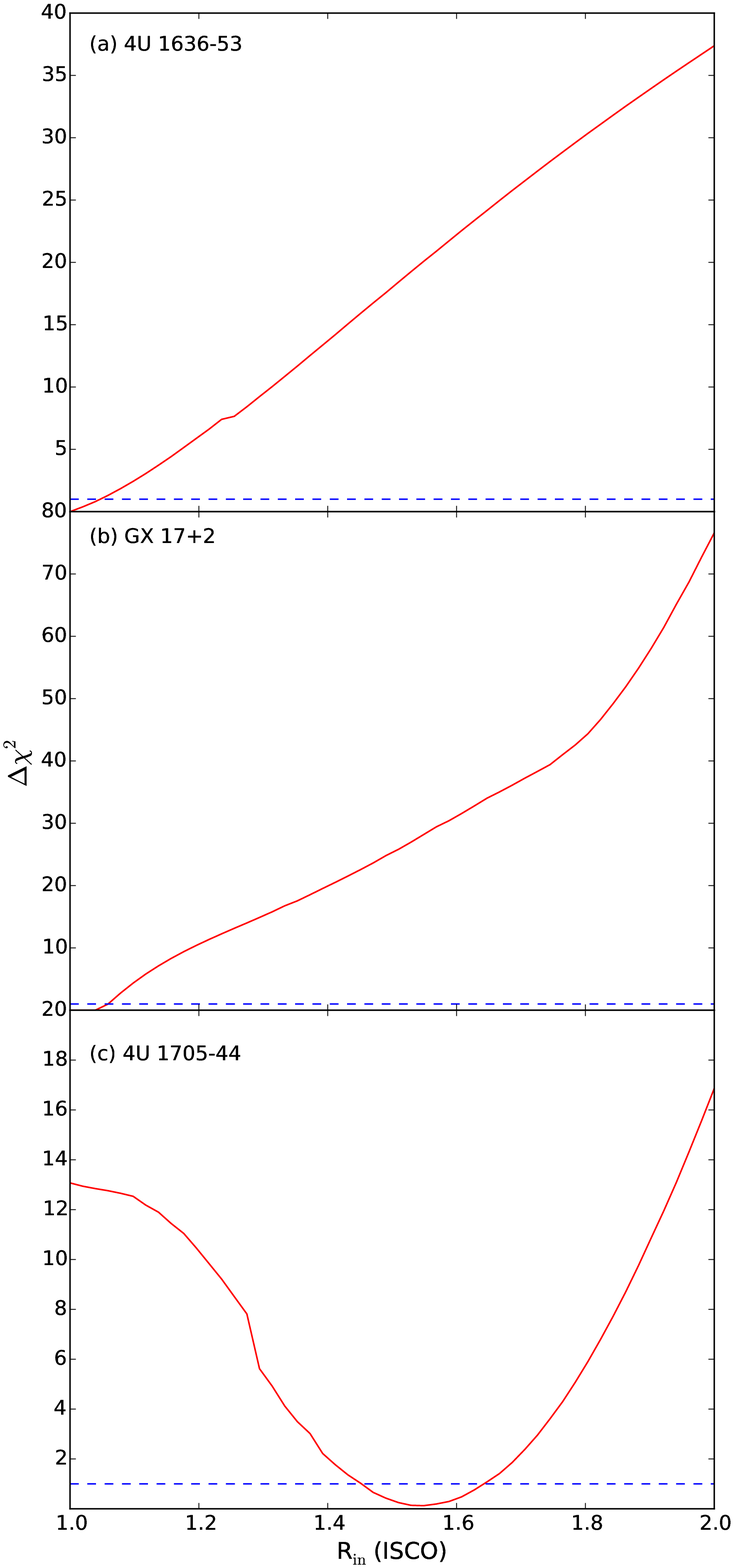}
\caption{The change in goodness-of-fit versus inner disk radius for the $\emph{NuSTAR}$ observations of 4U 1636-53, GX 17+2, and 4U 1705-44 taken over 50 evenly spaced steps generated with XSPEC \lq \lq steppar". The inner disk radius was held constant at each step while the other parameters were free to adjust. The blue dashed line shows the 68\% confidence level.
(a) 4U 1636-53 fit corresponding to the first column in Table 2. 
(b) GX 17+2 fit corresponding to the fifth column in Table 4. 
(c) 4U 1705-44 fit corresponding to the first column in Table 5.
}
\label{fig:steppar}
\end{figure}

\section{Discussion}
Using {\em NuSTAR}, we have taken a hard look at three well-known
neutron star X-ray binaries.  Our observations captured 4U 1636$-$53
in the hard state, while GX 17$+$2 and 4U 1705$-$44 were caught in
soft states.  Owing to $\emph{NuSTAR}$'s broad bandpass and its
ability to measure robust spectra at high flux levels, we are able to
constrain different properties of these sources through modeling of
reflection from their disks.  Different disk reflection spectra are
considered for each source in order to examine how inferred radii
depend on modeling assumptions.  For plausible combinations of neutron
star masses and dimensionless angular momenta, our results imply that
disks extend to an ISCO, that neutron stars are smaller than their
ISCO, and the results begin to place meaningful constraints on neutron
star radii.  In this section, we consider the results within the
context of the neutron star equation of state, implications for the
inner accretion flow onto neutron stars, and evaluate possible
systematic errors and avenues for improvement in future studies.

\subsection{Neutron Star Radius Constraints}

4U 1636-53 is found to have an inner disk radius of 1.00-1.03
  ISCO for $a_{*}=0.0$ and 1.02-1.14 ISCO for $a_{*}=0.3$ as
constrained from {\sc relxill} modeling of the reflection spectrum.  We
applied another reflection model, {\sc reflionx}, to test the
robustness of our measurement. The resulting fit gave a nearly
identical inner disk measurement.  For $a_{*}=0.3$ and 1.4
  $M_{\odot}$, $1.08\pm0.06$ ISCO translates to $10.8\pm0.6$ km. For
  the lower value of $a_{*}=0.0$ and 1.4 $M_{\odot}$, $1.03\pm0.03$
  ISCO translates to $12.4\pm0.4$ km. This small inner disk radius is
in agreement with previous measurements for this source that were
consistent with the ISCO (\citealt{pandel08}; \citealt{cackett10};
\citealt{sanna13}). The high inclination of 4U 1636-53 is consistent
with previous reflection studies (\citealt{frank87};
\citealt{casares06}; \citealt{cackett10}; \citealt{sanna13}).

Similarly, the inner disk radius in GX 17$+$2 is tightly constrained
to  1.00-1.02 ISCO for $a_{*}=0.0$ and 1.03-1.30 ISCO for
  $a_{*}=0.3$ by using the blackbody reflection model {\sc
  bbrefl}. We found a tighter constraint on the inner disk radius
  of 1.00-1.01 ISCO when we applied a {\sc reflionx} model that was
modified for a blackbody illuminating the disk. For $a_{*}=0.0$
  and 1.4 $M_{\odot}$, $1.00-1.02$ ISCO translates to $12.0-12.2$
  km. For the higher value of $a_{*}=0.3$ and 1.4 $M_{\odot}$,
  $1.03-1.30$ ISCO translates to $10.3-13$ km. The inclination was
  found to be $25^{\circ}-38^{\circ}$. \citet{cackett10} found
similarly small inner disk radii and low inclination for this system.

Thus, in the most sensitive and robust spectra of 4U 1636$-$53 and GX
17$+$2 yet obtained, the innermost extent of the accretion disk is
found to be close to the ISCO, with consequences for the neutron star
radii.  There may still be a boundary layer present on the surface of the NS, but in this case it would have to be quite small. Given that 1 ISCO corresponds to 12 km, for $a_{*}=0$ and 1.4 $M_{\odot}$, and using the fiducial neutron star radius of 10 km, the boundary layer would be about $\sim$2 km.  The inner disk radii are not strongly dependent upon the reflection
models that are utilized.  We therefore proceed to examine the
implications of the results for the EOS of ultradense matter in a more
generalized way.  Rather than assuming specific values $a =
cJ/GM^{2}$, we need to determine the likely range of $a$ given the
measured spin frequencies for these sources.

4U 1636-53 and GX 17$+$2 have known rotation frequencies: $\nu_{1636-53} =
518.0$ Hz \citep{Galloway08} and $\nu_{17+2} = 293.2$ Hz
\citep{wij97}, respectively. The total angular momentum, $J$, can be
obtained from the spin frequency assuming a reasonable range of mass
and radius for a neutron star and a solid sphere ($J=\frac{2}{5} M
R^{2} \omega$ where $\omega=2\pi\nu_{spin}$).  For 4U 1636$-$53, $\nu
= 518$~Hz then implies $0.09\pm0.05< a_{1636-53}< 0.44\pm0.23$, and $\nu
= 293.2$~Hz implies $0.05\pm0.04< a_{17+2}< 0.25\pm0.13$.  These
values assume a mass range of $1.3 \leq M_{\rm NS}/M_{\odot} \leq
2.1$, consistent with the range of masses that have been measured
directly (\citealt{jacoby05}; \citealt{Dem10}; \citealt{freire11};
\citealt{kiz13}).  The lower limit of the radius range was determined
by where causality approximately intersects the largest measure
NS. The upper limit of the radius range was limited by break up
($a_{*}=0.7$).

\begin{figure*}
\centering
\includegraphics[width=16.0cm]{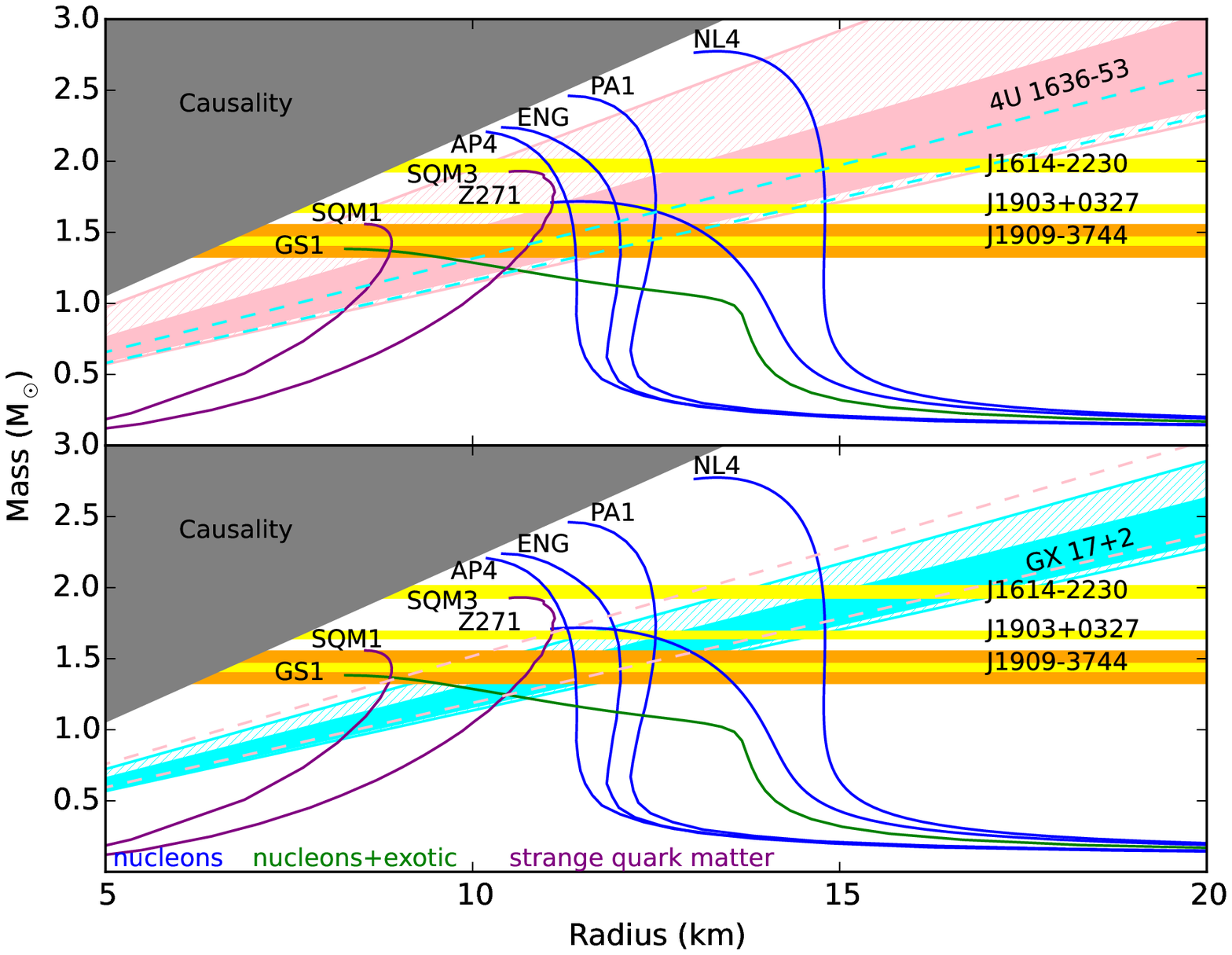}
\caption{Constraints on the cold, ultradense matter equation of state from Fe K$_{\alpha}$ reflection modeling to determine the inner disk radius, assuming that the stellar surface is truncating the disk. The gray region is excluded by causality (the speed of sound must be less than the speed of light). The curve labeled NL4 is from \citet{akmal98} and Z271 is from \citet{horowitz01}. All other mass-radius curves are labeled as in \citet{lattimer01}. The shaded regions for 4U 1636-53 and GX 17+2 correspond to the allowed values for mass and radius given a spin frequency of 518.0 Hz \citep{Galloway08} and 293.2 Hz \citep{wij97}, respectively. For reasonable values for mass and radius, a spin frequency of 518.0 Hz relates to a spin parameter of $0.09\pm0.05< a_{*}< 0.44\pm0.23$ and 293.2 Hz gives $0.05\pm0.04< a_{*}< 0.25\pm0.13$. The  hatched area represents the errors on the spin parameter. The dashed lines in each panel represent the solid area constraints from the NS in the other panel. The yellow horizontal lines are the measured masses for the NSs PSR J1614-2230 ($M=1.97\pm0.04 \ M_{\odot}$; \citealt{Dem10}), PSR 1903+0327 ($M=1.667\pm0.021\ M_{\odot}$; \citealt{freire11}) and PSR J1909-3744 ($M=1.438\pm0.024\ M_{\odot}$; \citealt{jacoby05}). The orange region represents mass range found for a NS in a double NS system ($M=1.33-1.55\ M_{\odot}$; see \citealt{kiz13} for a review).
}
\label{fig:eos}
\end{figure*}

The radius of the ISCO around a compact object in units of
gravitational radii is depends on its spin \citet{Bardeen72}; the
ranges above therefore enable a translation to gravitational radii and
then into kilometers.  Figure 4 plots these ranges in the mass versus
radius plane used to characterize the EOS.  Several equations of
states from \citet{akmal98}, \citet{lattimer01}, and
\citet{horowitz01} are also plotted, as well as known NS masses from
pulsar timing methods and binaries.  The regions allowed by our models
must be considered upper limits on the radius of neutron stars, since
the neutron star can be smaller than its ISCO.  Disk reflection is not
yet able to rule out plausible EOS; however, deeper X-ray spectra
and/or mass measurements in these systems can greatly reduce the
allowed regions in this mass--radius plane.

\subsection{Implications of Disk Truncation in 4U 1705$-$44}

In 4U 1636$-$53 and GX 17$+$2, the inner disk appears to extend to the
ISCO, and we cannot place any constraints on the magnetic field in
these sources.  In contrast, 4U 1705-44 has an inner disk radius of
1.46-1.64 ISCO for $a_{*}=0.0$ and $1.69-1.93$ ISCO for
  $a_{*}=0.3$. For a spin of 0.0 and stellar mass of 1.4 M$_{\odot}$,
  1.46-1.64 ISCO translates into 17.5-19.7 km. A spin of 0.3 and
  stellar mass of 1.4 M$_{\odot}$, 1.69-1.93 ISCO translates into
  16.9-19.3 km.

The similarity between the results of our fits and prior work suggests
that such modeling is converging on a relatively consistent set of
physical constraints.  A truncated disk has been indicated in 4U
1705$-$44 in several prior investigations (\citealt{disalvo09},
\citeyear{disalvo15}; \citealt{reis09b}; \citealt{Egron13};
\citealt{cackett10}, \citeyear{cackett12}). Our results closely match
those of \citet{reis09b}; that work reported the disk of 4U 1705$-$44
was truncated above the stellar surface with a gap of $\sim$3.5 km.
Our models find that the inclination of the inner disk is between
$24.0^{\circ}-26.1^{\circ}$; this is again largely consistent with
previous reflection studies ($20^{\circ}-50^{\circ}$,
\citealt{piraino07}; $\leq35^{\circ}$, \citealt{reis09b};
\citealt{dai10}; \citealt{cackett10}, \citeyear{cackett12};
\citealt{Egron13}). \citet{disalvo15} find a slightly larger
inclination than we do of $43^{\circ}\pm5^{\circ}$.

Disks around neutron stars can be truncated by a boundary layer, or by
magnetic pressure.  It is also possible that the inner disk may
evaporate at low accretion rates, qualitatively similar to the
expected truncation of disks around black holes at low mass accretion
rates.  Evidence of this may be seen in HETE J1900.1-2455
\citep{papitto13}.

In Section 3.3, we estimated that the boundary layer could push out to
1.2~ISCO based on the arguments in \citet{PS01}.  This is
smaller than the radius implied by most of our disk reflection models,
but the predicted extent of the boundary layer can be increased for
specific combinations of radiative efficiency and stellar spin.
\citet{dai10} estimated that the boundary layer in 4U 1705$-$44
extended out to $\sim 2\ R_{\mathrm{NS}}$ in the soft state. Our
estimate is consistent with this picture assuming a NS radius of 10
km.

If the neutron star magnetosphere is truncating the disk, we place an
upper limit on the magnetic field strength at the poles to be
$B\leq0.4-1.5\times10^{9}$ (see Section 3.3).  A recent study by
\citet{king16} also found a truncated disk surrounding the well-known
neutron star X-ray binary and pulsar Aql X-1.  An inner disk radius of
$R_{in}=15\pm3\ R_{g}=2.88\pm0.58$ ISCO was measured via disk
reflection. If the disk was not truncated by a boundary layer and
instead by the neutron star magnetic field, an upper limit of
$B<5\pm2\times10^{8}$ G results \citep{king16}.  Aql X-1 had a Type 1
X-ray burst during the observation, suggesting that some material was
still reaching the surface, like 4U 1705$-$44.

\subsection{Inner accretion flows and $\dot{M}$}

It is expected that the inner radius of an accretion disk is partly
set by the mass accretion rate through the disk (see, e.g.,
\citealt{accprop}).  Indeed, as noted above, several mechanisms might
truncate the inner accretion disk around a neutron star, but each
truncation mechanism becomes more effective at lower mass accretion
rates.  Indeed, the radial extent of the inner disk may be an
important factor in determining the phenomena manifested in different
parts of the ``Z'' and ``atoll'' tracks followed by many persistent
neutron star X-ray binaries, and even the position of the source along
such tracks.

The sensitive spectra that we have obtained with {\em NuSTAR} have
permitted particularly strong radius constraints.  Prior {\em NuSTAR}
observations of neutron stars have also measured inner disk radii.  It
is now pragmatic to consider what a modest collection of robust
spectral constraints implies about the evolution of the inner accretion
disk around neutron stars as a function of the mass accretion rate.
For consistency, we carefully replicated the models considered in
prior work within XSPEC, and extrapolated the models to a common
energy range (0.5--50~keV).  We then use the maximum distance to each
source to convert unabsorbed fluxes to luminosities, and divided by
an Eddington limit of $3.8\times10^{38}$ ergs s$^{-1}$
as per \citet{kuulkers03} to obtain the Eddington fraction.

Table 7 lists the key data that result from this exercise.  Figure 5
plots the inner disk radius versus Eddington fraction for a set of
eight neutron stars observed with {\em NuSTAR}.  Across almost two
orders of magnitude in Eddington fraction -- implying an equivalent
range in mass accretion rate if the efficiency is independent -- the
inner disk appears to remain very close to the ISCO.  The most obvious
exception to the overall trend is Aql X-1; however, this source is
known to be a pulsar and it may have a slightly higher magnetic field
than other sources in the sample.

\citet{cackett10} found similar results when looking at the inner disk
radius dependence on Eddington fraction for a sample of 10 neutron
stars that were observed with $\emph{XMM-Newton}$ and
$\emph{Suzaku}$. \citet{chiang16b} recently looked at $\emph{Suzaku}$
observations of Ser X-1 over a range of flux states and found that the
inner disk radius changes little between $0.2-0.6\ L_{\mathrm{Edd}}$.
Disk reflection studies undertaken with {\em NuSTAR} have the
advantages of a broad, continuous bandpass that enables better
characterization of the direct continuum and stronger constraints on the
total reflection spectrum, and spectroscopy that is not distorted by
photon pile-up (which can falsely skew the shape of the line:
\citealt{Ng10}; \citealt{miller10}).

\subsection{Potential systematic errors and modeling issues}
We have obtained spectra with very high statistical quality, and therefore spectral
fitting results with small statistical errors.  Systematic errors are
likely to be comparable or larger, and should be considered.  In practice,
the most important sources of systematic errors are difficult to
quantify, but they are important to mention.  

All measures of black hole spin that utilize the accretion disk, and
all limits on neutron star radii that utilize the disk (including disk
reflection and QPOs), assume that gas in the disk orbits as test
particles would orbit.  If real fluid disks push slightly
inward of the ISCO defined for test particles, it amounts to a
systematic error on such measurements.  There is no astrophysical test
of this assumption, but numerical simulations can potentially provide
some insights.  Explorations of disks around black holes suggest that
accretion disks do generally obey the test particle
ISCO \citep{reynolds08}.  Similar simulations have not been performed
in the presence of a neutron star, and the influence of a boundary
layer may also induce small changes.  New simulations can help to
address such systematics.

Apart from the reaction of the accretion disk to the potential, it is
possible that our assumption of a Kerr metric is itself a source of
systematic error.  The neutron star that we have studied may have a
small quadrupole moments, and this could change the effective
potential in which the disk orbits.  One treatment suggests that
systematic errors related to quadrupole-induced deviations from a pure
Kerr metric are $\leq$10\% even for a dimensionless spin parameter of
$a = 0.3$ \citep{SS98}.

The blurred reflection models themselves can potentially be sources of
systematic error.  Some ancillary parameters of the reflection models
require a note of caution.  We now turn to these issues.

For 4U 1636$-$53, both {\sc relxill} and {\sc relconv}$\times${\sc reflionx} return
small and formally consistent inner disk radii.  Different runs with
{\sc relconv}$\times${\sc bbrefl} and {\sc relconv}$\times${\sc reflionx} all return small
radii consistent with $R_{in} \leq 1.1$ ISCO in fits to GX 17$+$2.  In
some cases, the errors are formally different at the 1$\sigma$ level
of confidence, but the values are consistent over more conservative
ranges.  In these two cases, the models are in close agreement and
suggest that the radius values that emerge can be taken seriously.
For 4U 1705$-$44, a greater range of inner disk radii emerges with
different models and parameter selections.  The most recent model,
{\sc reflionx}, returns radii that are nearly consistent with the ISCO at
$1\sigma$; other models return values as large as $R_{in} = 1.8$ ISCO,
and the majority of models -- and the best overall fits -- measure a
truncated disk.  Deeper observations of 4U 1705$-$44 may be required
to obtain a more definitive picture of that source.

The emissivity profile of the accretion disk encodes the geometry of
the emitting source and space time of the innermost environment.  An
emissivity index of $q=3$ is expected for a point source emitter in a
(nearly) Schwarzschild spacetime, and different plausible geometries
for neutron stars (boundary layers, hot spots) appear to produce the
same emissivity profile (D. Wilkins, priv. comm.).  Both families of
reflection models prefer a flatter emissivity in fits to the spectrum
of 4U 1636$-$53.  This may be the result of a more extended corona in
the hard state.  Fits to the spectrum of GX 17$+$2 with {\sc bbrefl}
generally prefer a much steeper index, but fits with the more recent
{\sc reflionx} model are only slightly steeper than $q=3$.  This is
broadly consistent with the direct continuum, which suggests that a
boundary layer or hot spot is likely irradiating the disk.  The models
for 4U 1705$-$44 are all broadly consistent with $q=3$.

The abundance of iron affects the local {\it strength} of reflection
relative to the direct continuum, not the shape of the line.  It is
the {\it shape} of the disk reflection spectrum -- and particularly
the shape of the Fe K emission line -- that is used to infer the inner
radius of the accretion disk, and to thereby set an upper limit on the
radius of the neutron star.  The abundance of iron does not directly
affect our radius measurements.  However, it is interesting to
consider this parameter and whether or not it is accurately
determined.  Both families of reflection models prefer an iron
overabundance of 4.5--5.0 relative to solar values, in fits to 4U
1636$-$53.  It is unlikely that the abundance of iron is that high in
the low mass companion star in 4U 1636$-$53, but it is possible that
this measurement correctly describes the atmosphere of the accretion
disk.  There, the ionization structure may skew the relative
abundances of different elements to values that do not reflect the
overall abundances within the accretion flow.  Fits to the spectra of
GX 17$+$2 and 4U 1705$-$44 are consistent with iron abundances of
1.0--2.0 times solar values, but these fits also return lower
ionization parameters, potentially consistent with ionization
affecting abundance measurements.  It is also possible that the
enhanced abundances may the the result of effects in especially dense
gas.  In these cases, the abundance would increase to replicate the
continuum for a lower density disk allowed by the atomic data set
within the current reflection models \citep{garcia16}.

Finally, we simply note that the ionization parameters measured in
different fits to each source spectrum are comparable to the values
seen for other NS reflection studies, $2.3<log (\xi)<4.0$
(\citealt{cackett10}; \citealt{miller13}; \citealt{degenaar15}).

\section{Conclusions}
We have measured the inner disk radius for three different NSs that were observed with $\emph{NuSTAR}$. 4U 1636-53 and GX 17+2 have a small inner disk radius that is constrained to the ISCO. This value has proven to be model independent for 4U 1636-53 and suggest that the NS is smaller than their ISCO. Converting ISCO to km for a range of spin parameters and NS masses provides a range in which allowed theoretical EOSs must lie. 4U 1705-44 possesses a truncated disk which we used to explore the possibility of a magnetic field or a boundary layer.  We estimate the upper limit of the magnetic field surrounding the NS to be $B\leq0.4-1.5\times10^{9}$ G at the poles and depends on assumed conversion factor between disk and spherical accretion. We estimate the extent of a possible boundary layer out to $\sim1.2$ ISCO. 

Disk reflection has proven a valuable tool in determining properties of NSs, such as limits on the extent of the NS radius, boundary layer, and magnetic field strength. It provides another method to narrowing down the elusive EOS of ultradense, cold matter that makes up the NS. The advantage of this method being that the distance to the source is not needed and short observations can provide a clear look at Fe K$_{\alpha}$ emission. Furthermore, complementary mass estimates can yield further constraints to the EOS. 
\\
\\

\begin{figure} 
\centering
\includegraphics[width=9.2cm]{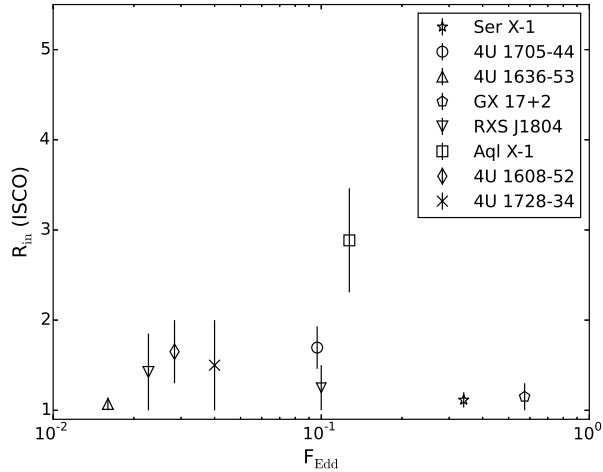}
\caption{Comparison of Eddington fraction and measured inner disk radii for NSs observed with $\emph{NuSTAR}$. Inner disk radius and Eddington fractions for 4U 1608-52, Ser X-1, Aql X-1, and 4U 1728-34 are obtained from \citet{degenaar15}, \citet{miller13}, \citet{king16}, and \citet{sleator16}, respectively. Values for RXS J1804 are taken from \citet{ludlam16} and \citet{degenaar16}. See Table 7 for inner disk radii and Eddington fractions. 
}
\label{fig:fedd}
\end{figure}

\begin{table}
\caption{NS Inner Disk Radii \& Eddington Fraction Observed with $\emph{NuSTAR}$ }
\label{tab:edd} 
\begin{center}
\begin{tabular}{lccc}
\hline
Source & $R_{in}$ (ISCO) & $F_{\mathrm{Edd}}$ & ref.\\
\hline
Ser X-1 & 1.03-1.20&0.34  & 1\\
4U 1705-44 & 1.46-1.93 & 0.10 &\\
4U 1636-53 & 1.00-1.14 & 0.01 &\\
GX 17+2 & 1.00-1.30 & 0.57 &\\
RXS J1804 & 1.00-1.85 & 0.02 & 2\\
& 1.0-1.5 & 0.10 & 3\\
Aql X-1 & 2.31-3.46 & 0.13 & 4\\
4U 1608-52 & 1.3-2.0 & 0.03 & 5\\
4U 1728-34 & 1.0-2.0 & 0.04 & 6\\

\hline
\end{tabular}

\medskip
Note.--- (1) \citealt{miller13}; (2) \citealt{ludlam16}; (3) \citealt{degenaar16}; (4) \citealt{king16}; (5) \citealt{degenaar15}; (6) \citealt{sleator16}.
\end{center}
\end{table}

This research has made use of the NuSTAR Data Analysis Software
(NuSTARDAS) jointly developed by the ASI Science Data Center (ASDC,
Italy) and the California Institute of Technology (Caltech, USA).  JMM
gratefully acknowledges support through the NuSTAR guest observer
program.  ND is supported via an NWO Vidi grant and an EU Marie Curie
Intra-European fellowship under contract
no. FP-PEOPLE-2013-IEF-627148.

\bibliographystyle{apj}
\bibliography{apj-jour,references}

\end{document}